\documentclass{ws-p8-50x6-00}

\begin{document}

\title{Possible Signatures of Quark-Hadron Phase 
Transitions inside Neutron Stars}

\author{Feng Ma}

\address{McDonald Observatory, the University of Texas at Austin, 
Austin, TX 78712-1083, USA\\Email: feng@astro.as.utexas.edu}
\maketitle

\abstracts{The spin-down power of an isolated neutron star
can drive its central density increase and overall structural
changes, and trigger a quark--hadron phase transition. 
A series of observational signatures may be seen
as a result of the phase transition, including pulsar 
spin-down and glitch behaviors, Soft Gamma 
Repeaters or Gamma-Ray Bursts. }

\section{Introduction}

The sky has long been a successful 
test ground and discovery site for fundamental physics. 
Examples include the discovery of Helium from the solar spectrum 
in 1868, evidence for gravity waves from binary pulsars~\cite{taylor79} in
1978, and more recently in 1985, 
the discovery of Bucky Balls (C$_{60}$), 
though not directly in the sky, 
 through experiments designed to simulate environments
near stars in favor of forming long chain carbon molecules~\cite{kroto85}. 

Quark--hadron phase transitions are being pursued 
in accelerators in the high temperature regime. 
Meanwhile, the densities inside neutron stars may be 
high enough to allow the existence of deconfined quark matter. 

\subsection{A Neutron Star with a Quark Core}

The structure of a neutron star with a quark core 
can be solved using standard 
equations of stellar structure~\cite{Shapiro}
together with the equations of state (EOSs) of quark matter
and normal neutron matter. 
The EOS of quark matter is 
much softer than that of neutron matter because of 
QCD asymptotic freedom. A neutron star containing
a quark core is thus more compact than a normal neutron star. 
Consequently, the star has a larger maximum spin frequency. 
However, submillisecond pulsars have not been discovered  
and hence quark stars have not been identified this way. 

\subsection{Rotating Relativistic Stars}

To solve the structure of a rotating 
relativistic star, Hartle
developed~\cite{Hartle1,Hartle2} a perturbation
solution based on the Schwarzschild metric of a static, spherically
symmetric object. 
Rotation distorts the star away from spherical symmetry,  and its
perturbed metric has the form 
\begin{equation}
 \displaystyle
  ds^2 = -e^{2\nu}dt^2+e^{2\lambda}dr^2+e^{2\mu}d\theta^2+
      e^{2\phi}(d\phi-\omega dt)^2 + O(\Omega^3).\label{eq:dif5}
\end{equation}
In this line element, $\omega$ is the angular frequency of the  star's fluid
in a local inertial frame and depends on the radial coordinate $r$. It
is related to $\Omega$, the rotational frequency seen by a
distant  observer.  
In this way, the  mass--central density
relation can be solved for  rotating compact stars~\cite{Hartle2,WG,Colpi}. 
We note that, the mass 
``increase", which is a main result in these works,  
represents the mass difference between two stars at different
angular velocities with the same central density. Hence, these theoretical 
results are  about star families relating  to  observations  
through  pulsar  statistics,  and are
difficult to be tested. 

\section{Predictions on Quark-Core Neutron Star Spin-Down Behavior}

\subsection{Another Way to Apply Hartle's Perturbation Method}

Instead of deriving mass--central density relations for a set 
of neutron stars, 
we have suggested to trace the evolution of an isolated 
neutron star during its spin-down process~\cite{maluo}. 
Associated with
the central density increase, neutron matter is 
continuously converted to quark matter.  
The overall structure and 
spin-down behaviors of the star are modified. 
In  Hartle's perturbation method,
the central density is fixed as an 
input parameter, then 
stellar masses are solved as a function of angular frequency. 
In principle, we can derive a set of equations 
parallel to those of Hartle, with a fixed mass as an input and central 
density as a perturbation, and solve the central density and 
stellar structure as a function of angular velocities.  
In practice, it is more convenient to apply Hartle's method
in a different way to
solve the change of central density of an isolated star. 

An isolated neutron star has approximately constant 
gravitational mass during its spin--down process.  
The binding energy of a  typical  neutron star
 $W{\sim}10^{53}$ erg, and the ratio  of  rotational  energy  to 
gravitational energy  is less than $0.1$ for the most rapidly  rotating 
neutron star~\cite{Colpi}. Since $M_{\odot}{\sim}10^{54}$ erg, the 
rotational energy is
only $1\%$ of the stellar mass. 
In calculating the stellar configurations
at different angular velocities, we need first to plot the mass--central
density  relation
at different angular velocities using Hartle's method as in 
previous work~\cite{Hartle2,WG,Colpi}. 
Under our approximation,  we cross these curves 
with a line of constant mass, and the corresponding central 
densities  are  those  of  a
rotating star at different angular velocities. 

\subsection{Pulsar Glitches}

The quarks inside the star are charged  particles and are coupled with the 
solid crust through magnetic fields. They rotate at a different velocity 
from superfluid neutrons. During the spin--down process, the central 
density and the quark core grow and so does the fractional moment of inertia of quark 
matter  as compared to the whole star ($I_q/I$). 
In a simple starquake model, the post--glitch behavior of a pulsar can be described 
approximately  as ~\cite{Shapiro},
\begin{equation}
  \displaystyle
            \Omega(t)=\Omega_{0}(t)+\Delta\Omega_0 [Qe^{-t/\tau}+1-Q],
             \label{eq:dif10}
\end{equation}
where $Q$ is the healing parameter of the pulsar. 
$Q=-\tau\Delta\dot\Omega(t{=}0)/\Delta\Omega$, and
$\tau=-\Delta{\Omega}/\Delta\ddot\Omega(t{=}0)$.
It has been shown that $Q{\propto}I_n/I$ (ref. 3), 
where $I=I_n+I_c$ is the total
moment of inertia of the star,  $I_n$
is the moment of inertia carried by neutral particles,
and $I_c$ is that of charged particles. 
Hence, a decreasing $Q$ over a long term may indicate continuous
conversion from neutron matter to quark matter inside the pulsar.

\subsection{Spin-Down History}

The pulsar spin-down can be described as~\cite{Shapiro}
\begin{equation}
 \displaystyle
\dot{\Omega}=-k{\Omega}^n, 
\end{equation}
and the braking index $n$ increases as the moment of inertia decreases, 
\begin{equation}
 \displaystyle
{\Delta}n/n \propto -{\Delta}I/I. 
\end{equation}
As a result, $|\dot{\Omega}|$ is larger because the 
quark-core neutron star has a smaller $I$ and thus 
is easier to brake. At the same
time, the star tends to  ``spin--up'' 
during the early stages of the phase transitions causing a higher
$\Omega$ than in the normal neutron star
case. Consequently, the ``characteristic age'' 
\begin{equation}
 \displaystyle
\tau=\Omega/2|\dot{\Omega}|=k^{-1}{\Omega}^{1-n}
\end{equation} 
may {\it underestimate} the true age of the neutron star
with a quark core~\cite{ma94}, while for a normal neutron star
$\tau$ is an upper limit.

\section{Onset of the Phase Transition: Catastrophic or Continuous?}

\subsection{Event Rate}

In the previous section we considered the neutron star with a 
developed quark core in it. What if the central density
of a normal neutron star happens to increase from sub-critical
density to critical density? Apparently the chance for this 
to happen is low, and we can estimate the 
event rate as~\cite{ma96}
\begin{equation}
 \displaystyle
    R \simeq 10^{-5} \left(\frac{P_i}{20{\rm\, ms}}\right)^{-2} \left(\frac{R_{\rm NS
}}{10^{-2}}\right) {\rm yr}^{-1} {\rm galaxy}^{-1},  \label{eq:rate}
\end{equation}
where $R_{\rm NS}$ is  the average neutron star birth rate 
in units of per year per galaxy and $P_i$ is the initial spin period. 
$R$ does not strongly  depend on the exact critical density.
The event rate is low because the 
central density increase is small if the initial spin is slow. 
However, equation (\ref{eq:rate}) suggest that these types of 
events  happen at a finite rate, especially when we have
millions of galaxies in our view. While the accretion
power in astrophysics has been stressed in many 
astronomical environments, equation (\ref{eq:rate}) suggest that 
the ``spin-down power'' is not negligible. The 
spin-down power can apply to any rotating systems with 
a critical point, not limited to quark-hadron 
phase transitions, but include Kion or Pion condensations
inside neutron stars. Even for a rotating white dwarf
the spin-down power alone may trigger 
a type Ia supernova as the Chandraseckhar mass limit 
depends on the rotational velocity. 

\subsection{Catastrophic}

If the phase transition is catastrophic, 
it may happen on a  time scale of seconds. 
In such a short time the star contracts from 
a radius of $\sim$15 km to  $\sim$10 km, 
associated with a sudden spin up. Most
importantly, the gravitational energy released 
\begin{equation}
 \displaystyle
    E \sim \frac{GM^2}{R}\left(\frac{{\Delta}R}{R}\right) \simeq
10^{53} ~ \left(\frac{{\Delta}R}{R}\right) ~ {\rm ergs}  
\end{equation}
may be large enough to power a cosmological gamma-ray burst (GRB)~\cite{ma96}.

\subsection{Non-Catastrophic}

Even though quark-hadron phase transition is
likely to be first order, it is different
from water--vapor phase transition in that 
the former has an additional freedom of 
whether a proton deconfines to $uud$ quarks
or a neutron becomes $udd$ quarks, which does not
conserve electric charges in the two phases although
the overall charge neutrality is conserved with the help
from leptons. Hence, the two phases are not necessarily separated
by gravity~\cite{glenn92}. In this case, the phase transition
may happen slowly on a time scale of $10^5$ years and 
the energy is released at an average  rate of
$10^{40}$ ergs s$^{-1}$. 
The majority of this energy is 
released via neutrino emission, 
and only a tiny fraction is used to heat the star up to a surface 
temperature of $3{\times}10^{6}$ K, 
yielding a soft X-ray luminosity of 
${\sim}10^{35}$ ergs s$^{-1}$. 
Note that it is 25 times more luminous than the sun, 
and it is not necessarily in a binary system~\cite{ma98}. 
While the fluid core of the star is contracting, 
the solid crust may have stress built up in it, and the
cracking of the crust can release bursts of energy~\cite{duncan92}.

\section{Observations}

In this section we confront theoretical predictions with observations. 
Radio pulsars constitute only a small fraction of all
neutron stars because of the beaming nature of their
radio emissions, and hence are not the best place
to look for the signatures of the phase transition. 
So far only 3 pulsars have reliably measured $n$. 
Five glitches for Crab pulsar, 
and 7 for Vela are recorded with measured $Q$. 
Many more are needed to see the continuous 
phase transition in radio pulsars.

\subsection{Soft Gamma Repeaters}

Soft Gamma Repeaters (SGRs) are X-ray transient sources 
(with $\sim$20 keV photons)
associated with young ($10^4$ yr) 
supernova remnants (SNRs). 
They are usually also quiescent X-ray emitters 
(with $kT{\sim}$1 keV, $L_{\rm X}{\sim}10^{35}$ ergs s$^{-1}$). 
So far 4 SGRs have been discovered in the Galaxy, 
and 1 in the Large Magellanic Cloud. 
Two SGRs show $\Omega/2|\dot{\Omega}|$ (${\sim}10^3$ yrs) 
$<$ SNR age (${\sim}10^4$ yrs)~\cite{kouv99}. 
Since many more Anomalous X-ray Pulsars (AXPs) share
some properties of SGRs, it seems that our 
prediction in previous section that we should be 
able to see one at a time in our Galaxy is an
underestimate. Note however that if the initial
spin of a neutron star is 2 ms rather than 20 ms (that of 
the Crab pulsar), we
get 100 times more event rate of phase transition.  
Fast initial spin has indeed been suggested to 
produce strong magnetic fields~\cite{duncan92} and 
slow down neutron stars much faster than Crab-type
pulsars. Radio astronomers may have been biased
toward Crab-type pulsars when picking up neutron stars.

\subsection{Gamma-Ray Bursts}

GRBs are energetic explosions with total 
energy $10^{51}-10^{54}$ erg (if isotropic), 
most of which is released with 100 keV to a few MeV 
photons. The time scales range from milliseconds to minutes. 
The event rate is estimated to be $10^{-5}$ yr$^{-1}$ galaxy$^{-1}$ if 
the emission is isotropic, and can be much higher if 
the emission is highly beamed. We note that 
the event rate in equation (\ref{eq:rate}) can still account for the GRBs
even if they are beamed, and the faster initial spins 
of the neutron stars naturally offer large angular 
momentum for the sake of beaming the electromagnetic 
radiations.

Gravity wave detectors such as LIGO may be able to observe 
gravity waves associated GRBs, and tell
which is the correct model. However, 
the gravity wave is emitted more or less isotropically 
while the electromagnetic radiation is likely to be highly
beamed (with a beaming factor 100--1000) in the 
angular momentum direction. Hence,  LIGO may need to 
accumulate 100--1000 events before  seeing a gravity wave signal
at the same time of a GRB. 

\section{Conclusions}

The accretion power in astrophysics has been stressed 
over decades, while the ``spin--down power'' is often 
neglected. Here we show that, in any astrophysical 
systems with angular momentum and a critical point, 
the spin--down power can drive the system to the critical
point at a small but finite probability. If the 
transition to critical state is energetic enough to
be seen from distant galaxies, we can observe these 
events regularly, as in the case for the signatures of quark
deconfinement phase transitions in radio pulsars, 
SGRs, and GRBs. 

\section*{Acknowledgments}

I thank Peter Williams for help with the manuscript.

\end{document}